\documentclass{ifacconf}
\usepackage{amsmath}
\usepackage{graphicx}      
\usepackage{natbib}        
\usepackage[all]{xy}
\usepackage{amssymb,amsfonts,ulem}
\usepackage{stackengine}
\usepackage{scalerel,stackrel}
\usepackage{color}

\newtheorem{theorem}{Theorem}
\newtheorem{proposition}[theorem]{Proposition}

\theoremstyle{definition}
\newtheorem{definition}[theorem]{Definition}
\newtheorem{example}{Example}

\begin{document}
\begin{frontmatter}

\title{Designing Poisson Integrators Through Machine Learning} 

\thanks[footnoteinfo]{The authors acknowledge financial support from the Spanish Ministry of Science and Innovation under grants PID2022-137909NB-C21, RED2022-134301-TD, the Severo Ochoa Programme for Centres of Excellence in R\&D (CEX2019-000904-S) and  BBVA Foundation via the project “Mathematical optimization for a more efficient, safer and decarbonized maritime transport”.}

\author[First]{Miguel Vaquero} 
\author[Second]{David Mart\'in de Diego} 
\author[Third]{Jorge Cort\'es}

\address[First]{IE University, 
   Segovia, 40001 Spain (e-mail:mvaquero@faculty.ie.edu).}
\address[Second]{Instituto de Ciencias Matem\'aticas ICMAT, 
  Madrid, 28049 (e-mail: david.martin@icmat.es)}
\address[Third]{University of California, San Diego, 
   9500 Gilman Dr, La Jolla, California, 92093-0411, (e-mail: cortes@ucsd.edu)}

\begin{abstract}                
This paper presents a general method to construct Poisson integrators, i.e., integrators that preserve the underlying Poisson geometry. We assume the Poisson manifold is integrable, meaning there is a known local symplectic groupoid for which the Poisson manifold serves as the set of units. Our constructions build upon the correspondence between Poisson diffeomorphisms and Lagrangian bisections, which allows us to reformulate the design of Poisson integrators as solutions to a certain PDE (Hamilton-Jacobi). The main novelty of this work is to understand the Hamilton-Jacobi PDE as an optimization problem, whose solution can be easily approximated using machine learning related techniques. This  research direction aligns with the current trend in the PDE and machine learning communities, as initiated by Physics-Informed Neural Networks, advocating for designs that combine both physical modeling (the Hamilton-Jacobi PDE) and data.
\end{abstract}

\begin{keyword}
Poisson geometry, symplectic geometry, geometric integrators, optimization, machine learning.
\end{keyword}

\end{frontmatter}

\section{Introduction}
Due to their persistent presence in science and engineering, Hamiltonian systems have been intensively studied for centuries. Special attention has been given to all the structures able to describe Hamiltonian systems, namely symplectic  and Poisson geometry. It is broadly recognized that geometry plays a pivotal role in the dynamical behavior of the aforementioned systems.  Nonetheless, since Hamiltonian systems can describe a wide array of systems in nature, they usually show a high level of complexity that hinders their complete understanding. This fact has led several communities to the design of algorithms that seek to produce accurate simulations of Hamiltonian systems as an enabling tool for the analysis of their dynamical behavior and properties.

To tackle this challenge, we follow here the geometric approach. The main observation is that numerical schemes sharing the same geometric properties as the original system usually enjoy better accuracy and more faithful qualitative description of the  system compared to non-geometric algorithms. This philosophy has been broadly exploited when the underlying geometry is symplectic, see for instance~\cite{serna,hairer}. 

Nonetheless, when the Hamiltonian system is described using the more general Poisson setting, the situation is  more subtle. See \cite{cosserat} and the references therein. This is mainly due to the fact that Poisson geometry is, in a  way, singular and irregular when compared to symplectic geometry. Moreover, to the best of the authors' knowledge, there are no general methods for obtaining Poisson integrators\footnote{By integrator we mean a numerical scheme designed to approximate the dynamics of a system. In this case, of a Hamiltonian vector field on a Poisson manifold.}, although some recent attempts include those in~\cite{cosserat}. In cases where the Poisson structure is linear and integrable, the references~\cite{Ge,Ge-Marsden,MMV,FLMV} provide methods to  generate them. These methods rely on the exploitation of the properties of the symplectic groupoid that integrates the Poisson manifold, or other symplectic realizations. Additionally, the same geometric structure was recently used to develop learning methods that conserve the underlying Poisson structure, as shown in~\cite{vaquero2023symmetry}. Also in~\cite{PeKeKa2023} the authors  design Poisson neural networks  to learn constant rank Poisson Hamiltonian systems   based on the Darboux-Lie theorem, finding first a coordinate change of coordinates that locally transforms the Poisson manifold as the product of a  symplectic manifold (with the canonical symplectic structure) and a manifold with the null Poisson bracket (see \cite{Weinstein} for a more general result).

In this paper, following the research direction initiated in~\cite{vaquero2023symmetry}, we propose a step further in the construction of Poisson integrators, combining differential geometry with machine learning techniques.  More precisely, we present the following contributions. {\it (1)} We introduce a general geometric setting for describing Poisson geometric integrators when the problem evolves on an integrable Poisson manifold. {\it (2)} We provide a method for approximating the Hamilton-Jacobi equation using machine learning-inspired techniques. {\it (3)} We illustrate our findings using the rigid body as an example.

\section{A Quick Review of Geometry}

\subsection{Basic Notions on Symplectic and Poisson Geometry}
In this section we introduce the main geometric structures used along the paper. We refer the reader to~\cite{Cra:21} for additional details.

\begin{definition}[{\it Symplectic Manifold}]
     A symplectic manifold is a pair $(M,\Omega)$, where $M$ is a manifold and $\Omega$ a non-degenerate closed two-form.
\end{definition}
\begin{example}
    The cotangent bundle of a manifold, say $T^*Q$, is endowed with the canonical symplectic form  $\omega_Q = -d\theta_Q$, where $\theta_Q$ is the Liouville one form \cite{AM87}. In local coordinates $(x^i,p_i)$,   $\omega_Q=dx^i \wedge dp_i$.  
\end{example}
Darboux's Theorem guarantees that given a symplectic manifold $(M, \Omega)$ there are local coordinates $(x^i, y_i)$ in which the coordinate representation of $\Omega$ is
$\Omega=dx^i\wedge dy_i$.
\begin{definition}[{\it Lagrangian Submanifold}]
    A Lagrangian submanifold $L$ of a symplectic manifold $(M,\Omega)$ is a submanifold of dimension half the dimension of $M$ and such that the restriction of $\Omega$ to $L$ is zero.
\end{definition}
\begin{example} Given a manifold $Q$, $\dim Q=n$, and its cotangent bundle $T^*Q$, any differentiable function $S: Q \rightarrow \mathbb{R}$ produces a Lagrangian submanifold in $T^*Q$ by just taking $graph(dS)\subset T^*Q$, this is called a horizontal Lagrangian submanifold. In canonical coordinates, $(x^i,p_i)$, this submanifold is just given by $(x^i, p_i=\displaystyle\frac{\partial S}{\partial x^i}(x))$, where $x\in Q$. The main
difficulty is to describe Lagrangian submanifolds which are not horizontal
(i.e.  the image of $dS$ ). It means that one can allow $S$ to be dependent not
only on the variables $(x^i)$, but on  mixed set of $n$-coordinates like
$(x^a,p_{n-a})$, $1\leq a\leq m<n$.  Of course, this  statement holds only locally (see \cite{arnold66}).   
  \end{example}

\begin{definition}[{\it Poisson Manifold}] A Poisson manifold is a manifold $P$ endowed with a bi-vector $\Pi$ satisfying $[\Pi,\Pi]=0$, where $[\cdot{},\cdot{}]$ is the Schouten bracket, see~\cite[Definition $2.3$]{Cra:21}.
\end{definition}
\begin{definition}[{\it Casimir function}] 
 A function $C: P\rightarrow {\mathbb R}$  
is called a Casimir function of a  Poisson manifold $(P, \Pi)$ if it verifies that  $\Pi( dC, dg)=0$ for all function $g: P \rightarrow \mathbb{R}$.
\end{definition}
Once we have set the basic objects of our constructions, we introduce the mappings conserving these geometric structures.

\begin{definition}[{\it Poisson Mappings}] Let $(P_1,\Pi_1)$ and $(P_2,\Pi_2)$ be two Poisson manifolds. Then, a mapping $f:P_1\rightarrow P_2$ is Poisson (resp. anti-Poisson) if $f_*\Pi_1 = \Pi_2$ (resp. $f_*\Pi_1 = -\Pi_2$). See~\cite[Definition $2.13$]{Cra:21} for the definition of the pushforward $f_*(\cdot{})$. 
\end{definition}
%
%

One of the main uses of Poisson manifolds is the description of Hamiltonian vector fields. 

\begin{definition}[{\it Hamiltonian Vector Field}] Given a Poisson manifold $(P, \Pi)$ and a function (Hamiltonian) $H:P \rightarrow \mathbb{R}$, we define the associated Hamiltonian vector field, $X_H$, as the unique vector field satisfying $X_H(g) = \Pi(dH,dg)$ for all functions $g: P \rightarrow \mathbb{R}$, see~\cite{Cra:21}.
\end{definition}

The {\it flow} of the Hamiltonian vector field $X_H$ is denoted by $\varphi^H_t$. A basic result in Poisson geometry states that $\varphi^H_t$ is a Poisson diffeomorphism for all fixed $t$.

\subsection{Poisson manifolds, Symplectic Realizations, and  \newline  Symplectic Groupoids}
Poisson manifolds are singular and difficult to deal with. A common approach in mathematics for analyzing complex structures is to seek ``desingularizations''. In other words, we aim to obtain a ``nicer'' geometric structure (symplectic in our case) on a manifold, endowed with a projection over the Poisson manifold under a Poisson mapping. This structure is called a {\it symplectic realization}, see~\cite[Ch. $6$]{Cra:21}. The main advantage of this approach is that it allows us to use the better understood and manageable  techniques from symplectic geometry. Among them,  Lagrangian submanifolds and their generating functions play an instrumental role here. 

The motivation for introducing {\it symplectic groupoids} is twofold: (1) The study of symplectic realizations naturally leads to symplectic groupoids, as described in~\cite{coste,Cra:21}. Symplectic groupoids constitute symplectic realizations when the set of units is a Poisson manifold. Moreover, the structural mappings $sou$ and $tar$ (described below) become Poisson and anti-Poisson mappings. Remarkably, $sou$ and $tar$ exhibit a dual pair structure, as detailed~ in \cite[Ch $14$]{Cra:21} and~\cite[Section III]{coste}. (2) Lagrangian submanifolds in symplectic groupoids induce Poisson diffeomorphisms in the set of units (see Theorem~\ref{CDW} below).

\begin{definition}[{\it Groupoid}] A \textit{groupoid} $(G)$   consists of the following components:
\begin{enumerate}
    \item Two Sets: 
    \begin{itemize}
        \item A set of objects (or units), denoted ($G_0$).
        \item A set of arrows (or morphisms), denoted ($G_1$).
    \end{itemize}

    \item Source and Target Maps: 
    \begin{itemize}
        \item A source map  $sou: G_1 \rightarrow G_0$ assigning to each arrow an object called its source.
        \item A target map $tar: G_1 \rightarrow G_0$ assigning to each arrow an object called its target.
    \end{itemize}

    \item Composition Law: A partially defined composition (or multiplication) operation \( m: G_1 \times_{G_0} G_1 \rightarrow G_1 \), where \( G_1 \times_{G_0} G_1 \) is the set of all pairs of arrows \((g, h)\) such that the target of \( h \) is the source of \( g \) (i.e.,  $tar(h) =  sou(g)$). The result of the composition \( m(g, h) \) is an arrow from the source of \( h \) to the target of \( g \).

    \item Associativity: The composition of arrows is associative where defined. That is, for any three arrows \( f, g, h \) in \( G_1 \), if \( f \cdot (g \cdot h) \) and \( (f \cdot g) \cdot h \) are defined, then \( f \cdot (g \cdot h) = (f \cdot g) \cdot h \).
    \

    \item Identity Elements: For each object \( x \) in \( G_0 \), there exists an identity arrow \( e_x \) in \( G_1 \) such that  $sou(e_x) = x$,  $tar(e_x) = x$, and for any arrow \( g \) with $sou(g) = x$ or $tar(g) = x$, the compositions $e_x \cdot g$ and $g \cdot e_x$ are defined and equal to $g$.

    \item Inverses: For each arrow \( g \) in \( G_1 \), there exists an inverse arrow \( g^{-1} \) in \( G_1 \) such that $sou(g^{-1}) = tar(g)$, $tar(g^{-1}) = sou(g)$, and the compositions \( g \cdot g^{-1} \) and \( g^{-1} \cdot g \) are defined and equal to the identity arrows of $tar(g)$ and  $sou(g)$, respectively.
 
\end{enumerate}
\end{definition}
\begin{definition}[Symplectic Groupoid]
A symplectic grou\-poid is a grou\-poid \( G \) equipped with a symplectic structure (a non-degenerate, closed 2-form) on its space of arrows \( G_1 \), such that the graph of the multiplication map \( m: G_1 \times_{G_0} G_1 \to G_1 \) is a Lagrangian submanifold of \( G_1 \times G_1 \times \overline{G_1} \) (here \( \overline{G_1} \) denotes \( G_1 \) with  minus the given symplectic structure).   
\end{definition}

When there is no room for confusion, we identify the set of arrows $G_1$ and the groupoid itself, $G$. Although there are known geometric obstructions to obtaining global symplectic groupoids whose set of units is a given Poisson manifold,  it is possible to obtain a local symplectic groupoid, as discussed in~\cite{Cabrera}. We would like to emphasize that local existence is sufficient for our constructions, as they are inherently local in nature.
\begin{example}
Let $\mathfrak{h}$ a Lie algebra with Lie bracket $[\; ,\, ]$ and $H$ a Lie group integrating it. Consider the Poisson manifold $(G_0={\mathfrak h}^*, \Pi_{LP})$ where $\Pi_{LP}$ is the induced Lie-Poisson structure.
  \[\Pi_{LP}(dg_1,dg_2)(\mu)=-\mu([dg_1(\mu),dg_2(\mu)])\]
where $g_1, \ g_2: \ \mathfrak{h}^*\rightarrow\mathbb{R}$ and $\mu \in \mathfrak{h}^*$.
 In this particular case, we can take $G_1$ as $T^*H$  where the source and target maps are given by 
\begin{equation}\label{liealgebra}
\begin{array}{rccl}
sou:& T^*H&\longrightarrow &  \mathfrak{h}^* \\ 
& (\mu_h) &\longmapsto &sou(\mu_h)=J_L(\mu_h) ,\\ \noalign{\bigskip}
tar:& T^*H&\longrightarrow &  \mathfrak{h}^* \\ 
& (\mu_h) &\longmapsto &tau(\mu_h)=J_R(\mu_h).
 \end{array}
\end{equation}
where 
\[
\langle J_L(\mu_h),\xi\rangle=\langle \mu_h,T_e {\mathcal R}_h(\xi)\rangle\; ,\quad 
 \langle J_R(\mu_h),\xi\rangle=\langle \mu_h,T_e {\mathcal L}_h(\xi)\rangle
\] for all  $\xi\in {\mathfrak h}$. Here,  ${\mathcal R}_h$ and ${\mathcal L}_h$ denote the right and left translations on the Lie group $H$, respectively.
See more details and examples in \cite{Marle,FLMV}.
\end{example}

\section{Lagrangian Bisections and Poisson Diffeomorphisms: The Hamilton-Jacobi Equation}

Let $(P\equiv G_0,\Pi)$ be a given Poisson manifold and assume there exists a symplectic groupoid, $G$, having $G_0$ as its set of units. This assumption is always valid locally, so there is no loss of generality. Let $H:G_0\rightarrow \mathbb{R}$ be a Hamiltonian. Our goal is to design methods to accurately approximate $\varphi^H_t$. Towards this end, the stepping stone of our constructions is the following observation.
\begin{quote}
\textbf{Key Observation:}  Poisson diffeomorphisms in $G_0$, like $\varphi^H_t$, can be described through Lagrangian submanifolds in the symplectic groupoid $G$.     
\end{quote}
In this context, the next result is at the core of our construction of Poisson integrators.

\begin{theorem}[\cite{coste}]\label{CDW} Let $G$ be a symplectic groupoid. Let $L$ be a
  Lagrangian submanifold of $G$ such that $sou_{|L}: L\rightarrow G_0$ is a (local)
  diffeomorphism (a Lagrangian bisection). Then
\begin{enumerate}
\item $tar_{|L}:L\rightarrow G_0$ is a (local) diffeomorphism as well.
\item The mapping
  $\hat{L}= sou \circ tar_{|L}^{-1}:G_0\rightarrow G_0$ 
  is a (local) Poisson diffeomorphism.
\end{enumerate}
\end{theorem}

Motivated by this result, a natural question is how to characterize and find the Lagrangian submanifold $L$, inducing the transformation $\varphi^H_t$. The answer leads directly to a geometric version of the Hamilton-Jacobi equation, as discussed in~\cite{FLMV}. Due to space constraints, we provide here  only a local description of the Hamilton-Jacobi equation. Consider a symplectic groupoid $({G},\Omega)$ with Darboux coordinates, $(x^i,p_i)$ such that $\Omega$ reads as $dx^i\wedge dp_i$. We examine the extended space $\mathbb{R}^2\times {G}$ with coordinates $(t,p_t,x^i,p_i)$, where we think of $p_t$ as the dual variable to $t$. In this context, we consider the $2$-form with local expression $dt\wedge dp_t + dx^i\wedge dp_i$, which turns $\mathbb{R}^2\times {G}$ into a symplectic manifold. Below, we define a generalized solution of the Hamilton-Jacobi equation.

\begin{definition}
A Lagrangian submanifold $L$ in $\mathbb{R}^2\times {G}$ is a generalized solution to the Hamilton-Jacobi equation if
\begin{equation}\label{genHJ}
(p_t + H\circ sou)_{|L} = \text{constant}.  \tag{Gen. HJ}
\end{equation}
\end{definition}

When a generating function, say $S$, is used to describe the Lagrangian submanifold $L$, then equation~\eqref{genHJ} takes the usual PDE form. For simplicity in this paper, we only consider the case where the generating function depends on $(t,p_i)$. This approach is applicable to the Lie-Poisson case treated below. Then, the Hamilton-Jacobi equation reads as follows
\begin{equation}\label{HJ}
  \displaystyle\frac{\partial S}{\partial t}(t,p_i) + H( sou(\frac{\partial S}{\partial p}(t,p_i),p_i)) = \text{constant}.
\end{equation}

\begin{proposition}[\cite{FLMV}] The Lagrangian submanifold satisfying the Hamilton-Jacobi equation induces the Hamiltonian flow, $\varphi^H_t$, up to an initial condition. That is, we can recover $\varphi_t^H\circ f$ where $f$ is a Poisson diffeomorphism. If $S(0,p_i)$ induces the identity on $P$, then we recover exactly $\varphi_t^H$.
\end{proposition}
%
%

 \subsection{Approximating the Hamilton-Jacobi Equation} 
Since obtaining an analytical solution to the Hamilton-Jacobi equation is challenging, approximation methods must be considered. A classical approach involves obtaining a Taylor expansion of the generation function $S$ in the $t$-variable and deriving a recurrence relationship that enables the determination of a solution to a desired arbitrary high order (see~\cite{FLMV}). The main contribution of this paper is to present an alternative strategy based on recent developments and tools in machine learning, which involves viewing the PDE as an optimization problem.  To do this, we consider a neural network with weights $W$ parametrizing candidate functions, $S(t,p_i;W)$. Then, we select a set of points $\{(p_i)_j\}_{1\leq j\leq N}$, where the Hamilton-Jacobi equation is evaluated. These points can be obtained using different means: for instance, through uniform sampling in the region where we aim to build the approximation to the Hamilton-Jacobi equation.
%
%
Then, we enforce the Hamilton-Jacobi equation at the points $(p)_{1\leq j\leq N}$ through the minimization of the mean-square error loss (or any other loss function), and consider the problem:
{\small
\begin{equation}\label{HJML}
    \min\limits_{W}\frac{1}{N}\sum_{j}\left(\displaystyle\frac{\partial S}{\partial t}(t,(p_i)_j;W) + H(sou(\frac{\partial S}{\partial p}(t,(p_i)_j;W),(p_i)_j))\right)^2.
\end{equation}}
This problem penalizes scenarios where the Hamilton-Jacobi equation is not satisfied. To approximately solve it, we can use the recent advancements in machine learning, including backpropagation to compute gradients and efficient optimization algorithms such as SGD, Nesterov, ADAM and others. See~\cite{GoodBengCour16} for a detailed description of these topics.
%
%

\section{Numerical Simulations}
\begin{figure}
    \centering
    \includegraphics[width = 7cm]{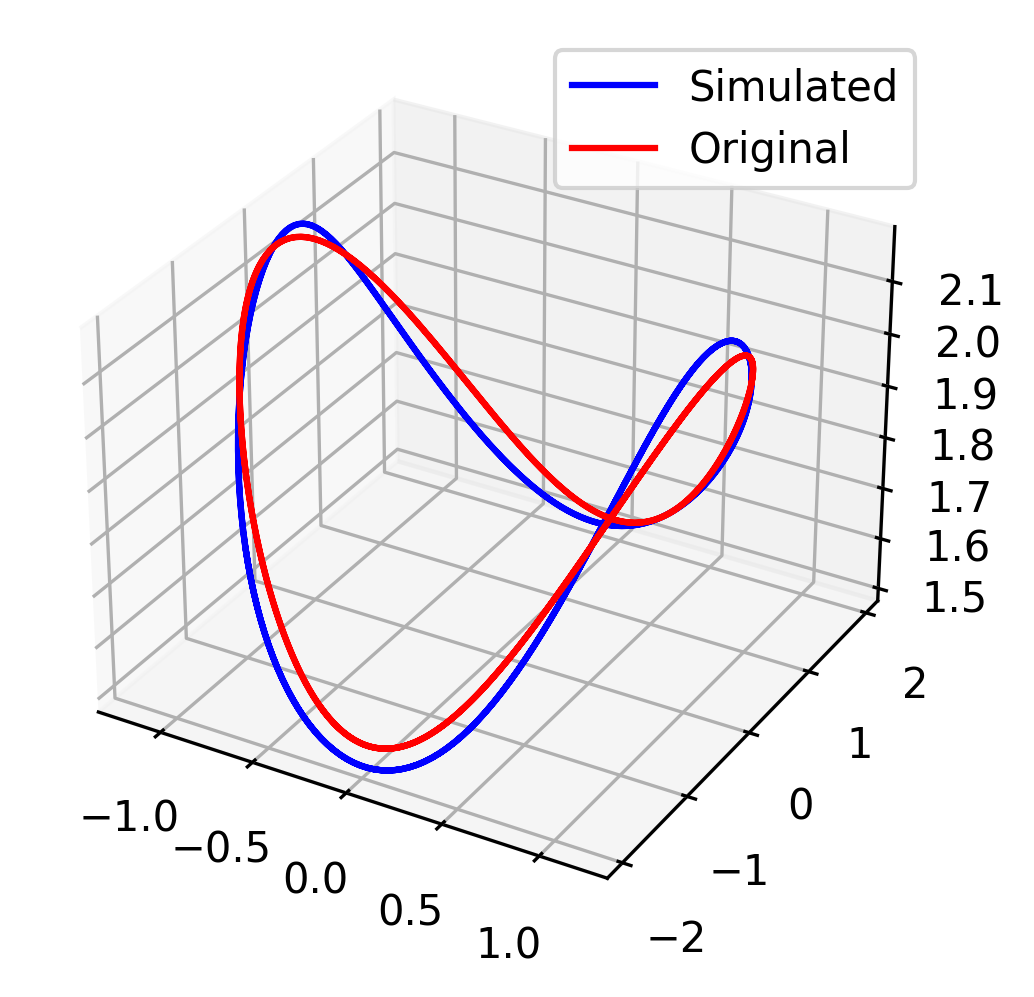}
    \includegraphics[width = 7cm]{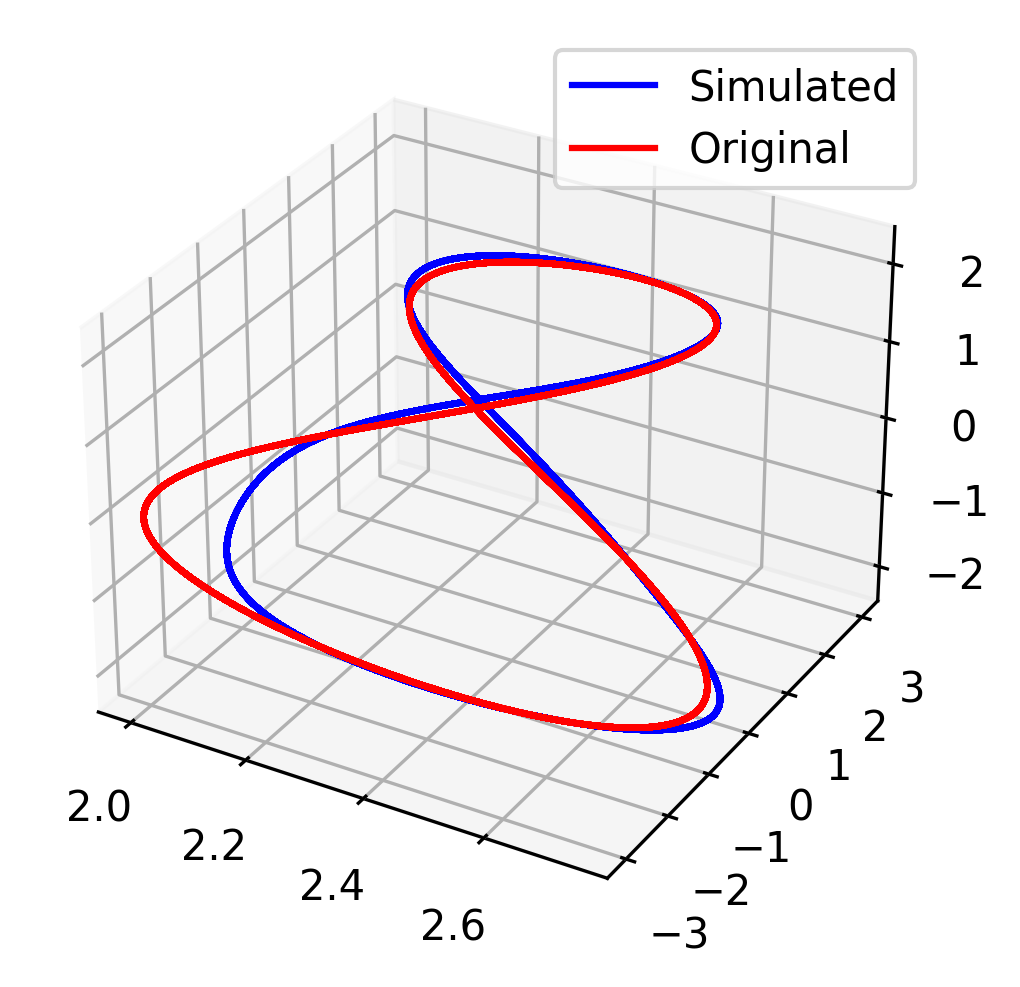}
    \caption{ Comparison of the simulated dynamics (blue) and the real trajectory (red) when the initial condition are the points $(1,1,2)$ {\it (top plot)} and $(3,2,0)$ {\it (bottom plot)}. 
    }
    \label{fig:2}
    \end{figure}

\begin{figure}
    \centering
\includegraphics[width = 7cm]{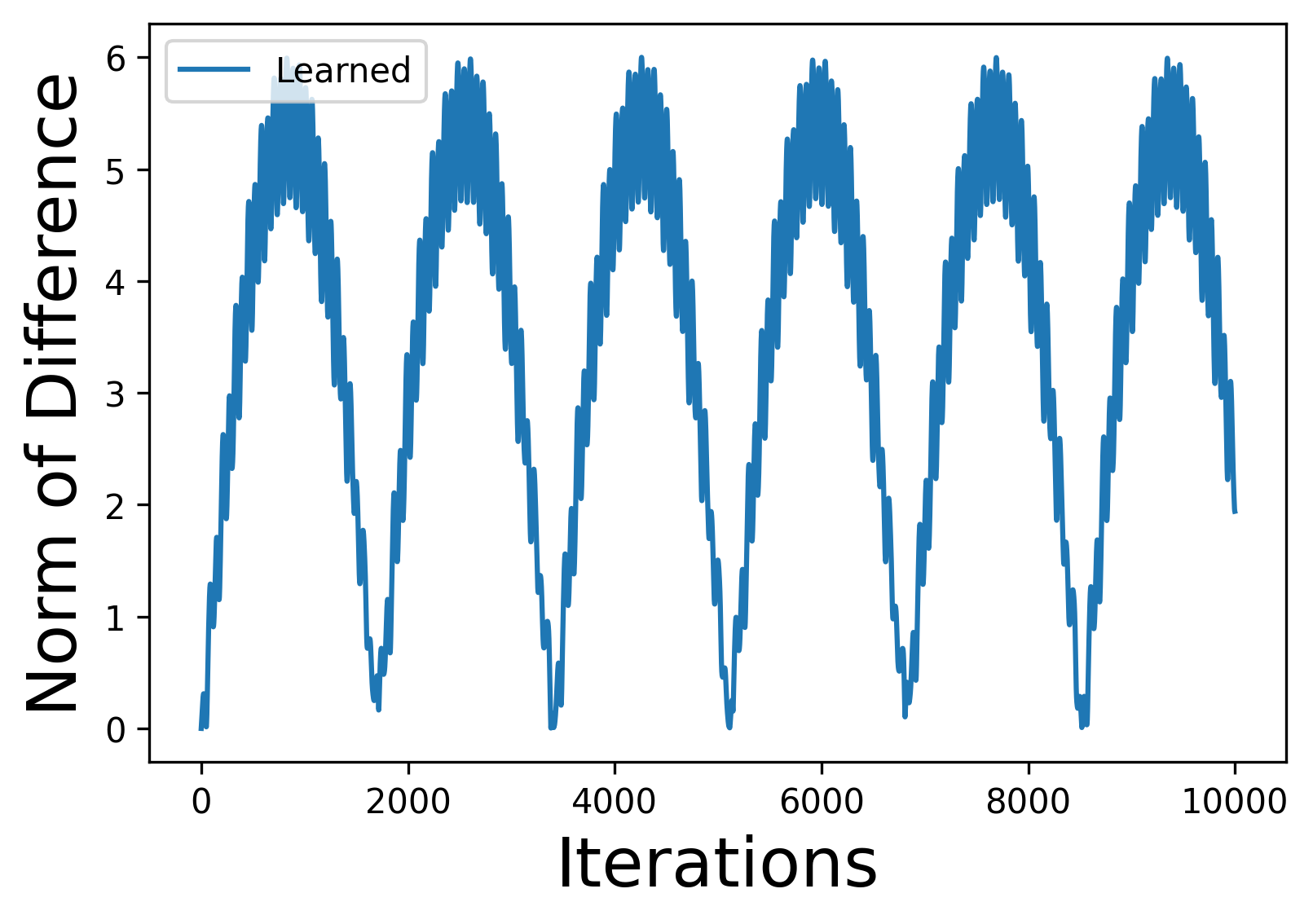}
    \includegraphics[width = 7cm]{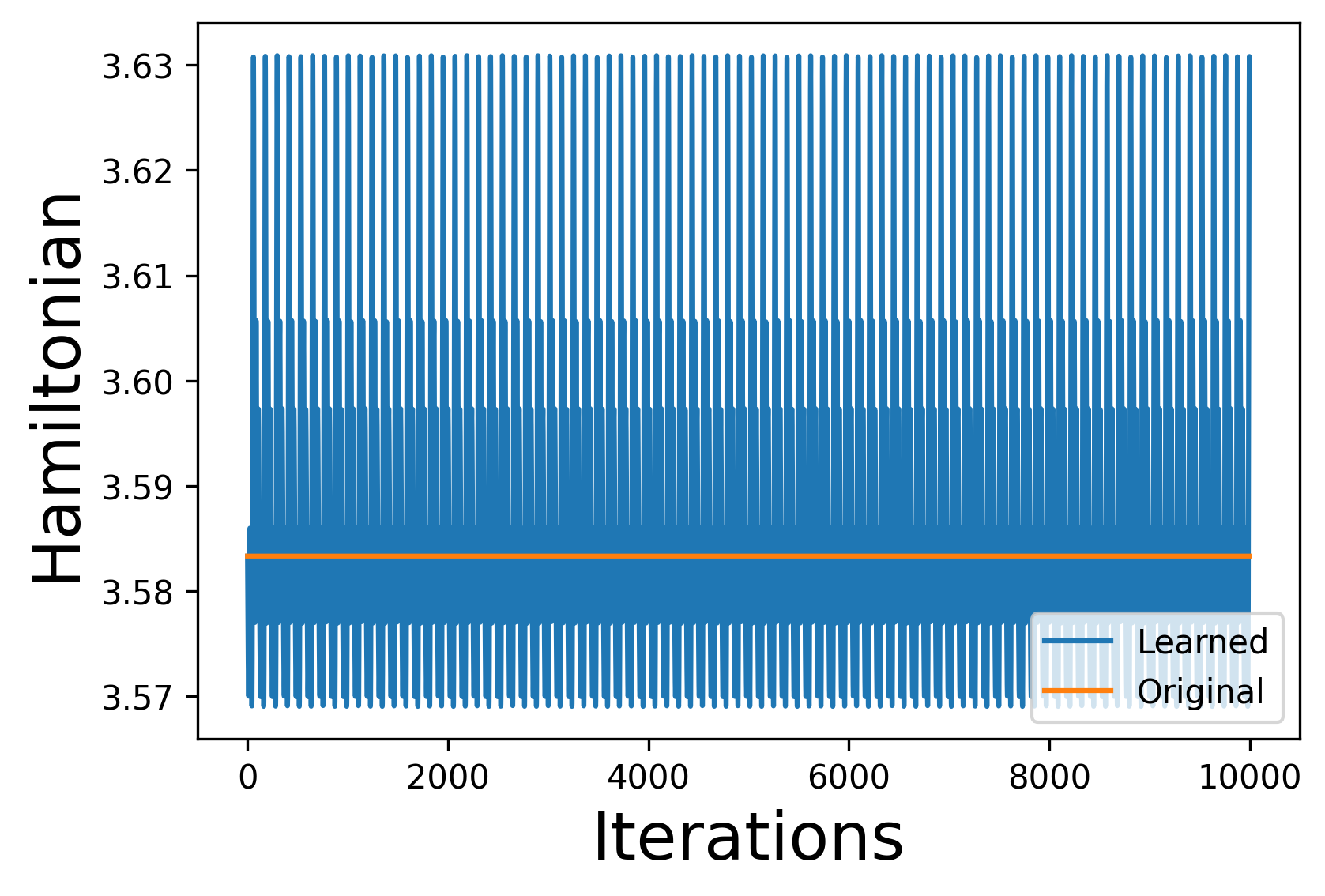}
    \includegraphics[width = 7cm]{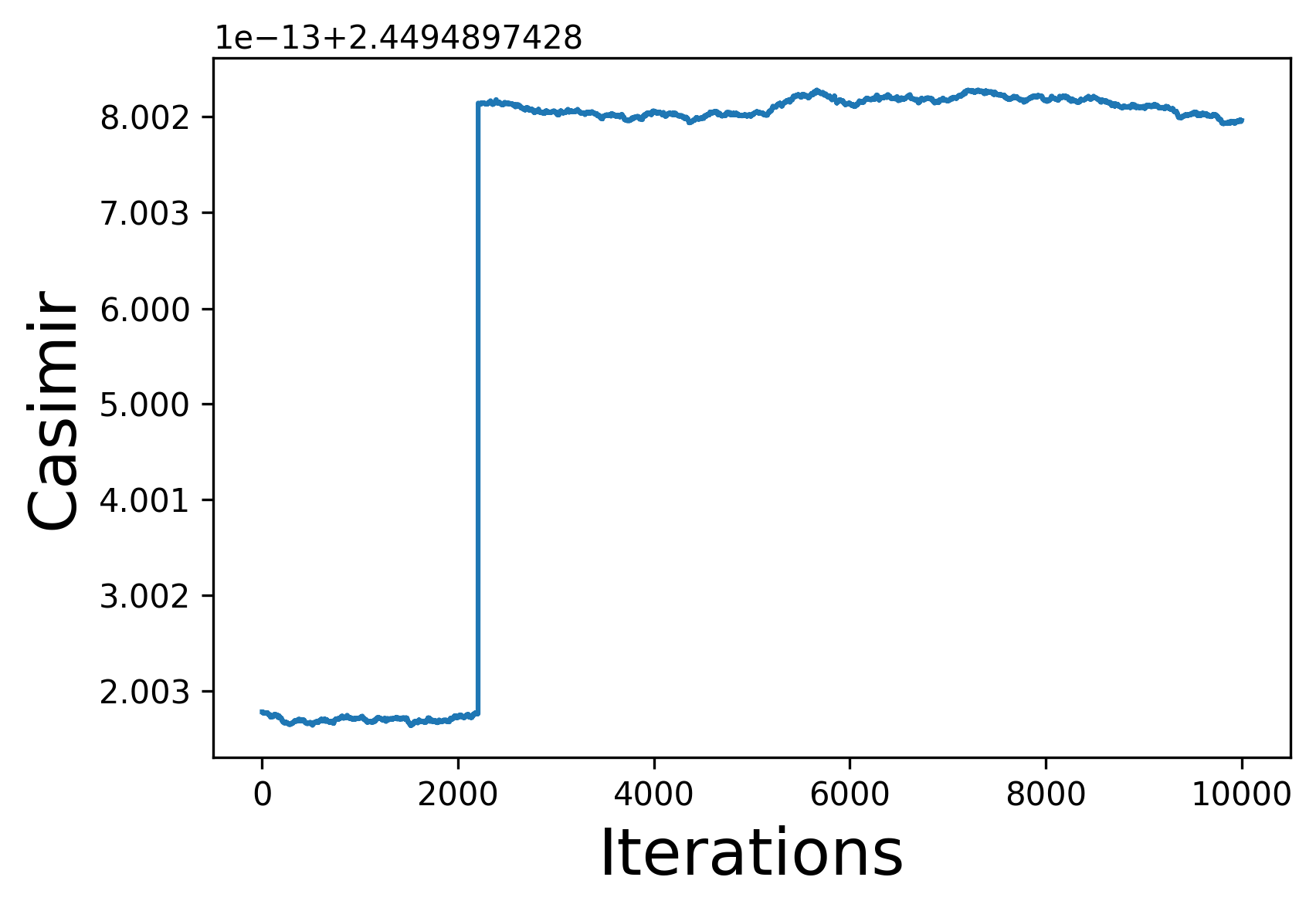}  
     \caption{{\it Top:} Evolution of the norm of the difference between the simulated and the real trajectories for $10,000$ iterations with stepsize $0.1$. The difference stays bounded due to the conservation of the underlying geometry. {\it Middle: } Evolution of the Hamiltonian when evaluated on the simulated trajectory. The Hamiltonian is almost conserved even for very long trajectories. {\it Bottom: } Evolution  of the Casimir. The Casimir function is conserved up to rounding error.}
    \label{fig:3}
    \end{figure}

In this section, we illustrate our findings by applying the described methodology to the rigid body, a benchmark system of Poisson dynamics. We consider the usual Lie-Poisson structure on $\mathfrak{so}^*(3)$, as discussed in~\cite{marsden3}, along with the Hamiltonian
\begin{align*}
    H = \frac{1}{2}\left(\frac{x^2}{1.5} + \frac{y^2}{2} + \frac{z^2}{2.5}\right) .
\end{align*}
It can be shown that $T^*SO(3)$ is a symplectic groupoid  having $\mathfrak{so}^*(3)$ as its set of units, and therefore constitutes a symplectic realization
%
%
(see~\cite{FLMV}). We uniformly sampled $80,000$ points around $(0,0,0)$, each coordinate varying between $-3$ and $3$ and used ADAM  ($10,000$ iterations with a learning rate of $10^{-4}$) to approximately solve~\eqref{HJML}. We employed a feed-forward neural network of four layers with $500-250-250-250$ neurons and activation function $\tanh$, using Pytorch for Python. We used uniform Xavier initialization for the weights.  Two different initial conditions were chosen to showcase the outcomes when the obtained integrators are used, as seen in Fig.~\ref{fig:2}. The qualitative behaviour of the simulated trajectories is very similar to that of the original system. In both cases, we observe a nice conservation of the Hamiltonian throughout the simulated trajectories, even for very long simulations. As expected, the Casimir is conserved in both cases up to round-off error. The evolution of the Hamiltonian, Casimir and the difference from the original dynamics for a long trajectory ($10,000$ iterations with stepsize $0.1$) is illustrated in Figure~\ref{fig:3}.

%

\section{Conclusions and Future Work}
In this paper, we delved into the geometric setting presented previously in~\cite{vaquero2023symmetry}.
%
%
The framework introduced possesses several properties that make it compelling for the design of Poisson integrators: it fully respects the underlying Poisson geometry and it is highly flexible, making it amenable to various modifications and improvements. Here, we describe a couple of research directions that we plan to pursue.

\emph{General Poisson Structures:}
When the symplectic groupoid is not readily available, several means can be used to approximate it. Two main constructions~\citep{Cabrera} have been proposed to produce local symplectic groupoids that can integrate (locally) any Poisson manifold. An ongoing research direction aims to exploit these constructions to extend the presented framework to any Poisson manifold.

\emph{Combination with Data:}
In some situations, trajectories of the system are available. These trajectories might have been obtained through other (geometric or non-geometric) integrators or might correspond to actual measurements. In these cases, we envision a blended approach that combines both the data and the Hamilton-Jacobi equation.  In this fashion, we follow the spirit of PINNS as described by~\cite{RPK2019}, designing integrators though finding the Lagrangian submanifold that minimizes an objective function of the type $loss = loss_{HJ} + loss_{data}$. In this objective function $loss_{HJ}$ ensures that the Hamilton-Jacobi equations is satisfied to a certain degree, following the same pattern as in~\eqref{HJML}. The term $loss_{data}$ would make the Lagrangian submanifold $L$ induce a Poisson transformation that matches the given data. 


\bibliography{References.bib}








\end{document}